\documentclass[12pt,preprint]{aastex}
\usepackage[]{times, graphicx}
\begin{document}

\shorttitle{Absorption of Transition Region Moss Emission}
\shortauthors{De Pontieu, Hansteen, McIntosh \& Patsourakos} 
\title{Estimating the Chromospheric Absorption of Transition Region Moss Emission} 
\author{Bart De Pontieu\altaffilmark{1,6},
  Viggo H. Hansteen\altaffilmark{2,1}, Scott W.
  McIntosh\altaffilmark{3}, Spiros Patsourakos\altaffilmark{4,5}} 
\altaffiltext{1}{Lockheed Martin Solar and
  Astrophysics Lab, 3251 Hanover St., Org. ADBS, Bldg. 252, Palo Alto,
  CA 94304} 
\altaffiltext{2}{Institute of Theoretical Astrophysics,
  University of Oslo, Blindern, Oslo 315, Norway}
\altaffiltext{3}{High Altitude Observatory, National Center for
  Atmospheric Research,P.O. Box 3000, Boulder, CO 80307}
\altaffiltext{4}{Naval Research Laboratory, Space Science Division, Washington, DC 20375, USA}
\altaffiltext{5}{Center for Earth Observing and Space Research, George Mason University, Fairfax, VA 22030, USA}
\altaffiltext{6}{Electronic address: bdp@lmsal.com}

\begin{abstract}
Many models for coronal loops have difficulty explaining the observed EUV brightness of the transition region, which is often significantly less than theoretical models predict. This discrepancy has been addressed by a variety of approaches including filling factors and time-dependent heating, with varying degrees of success. Here we focus on an effect that has been ignored so far: the absorption of EUV light with wavelengths below 912 \AA\ by the resonance continua of neutral hydrogen and helium. Such absorption is expected to occur in the low-lying transition region of hot, active region loops, that is co-located with cool chromospheric features and called ``moss'' as a result of the reticulated appearance resulting from the absorption. We use co-temporal and co-spatial spectroheliograms obtained with SOHO/SUMER and Hinode/EIS of Fe~{\sc xii} 1242\AA, 195 \AA\ and 186.88 \AA, and compare the density determination from the 186/195 \AA\ line ratio to that resulting from the 195/1242 \AA\ line ratio. We find that while coronal loops have compatible density values from these two line pairs, upper transition region moss has conflicting density determinations. This discrepancy can be resolved by taking into account significant absorption of 195 \AA\ emission caused by the chromospheric inclusions in the moss. We find that the amount of absorption is generally of order a factor of 2. We compare to numerical models and show that the observed effect is well reproduced by 3D radiative MHD models of the transition region and corona. We use STEREO A/B data of the same active region and find that increased angles between line-of-sight and local vertical cause additional absorption. Our determination of the amount of chromospheric absorption of TR emission can be used to better constrain coronal heating models. 
\end{abstract}



\section{Introduction}
The nature of the transition region continues to be the subject of
significant debate. As part of the interface between the photosphere
and corona, it plays an important role in the transport and deposition
of the non-thermal energy believed to power the heating of both
chromosphere and corona. It is now well established that the interface
between the photosphere and corona does not resemble the view suggested by
hydrostatic, spherically symmetrical and time-averaged models. Instead,
the chromosphere is a highly dynamic environment \citep{Rutten2006} with jets, such as
fibrils, spicules and mottles \citep{Hansteen2006,DePontieu2007,Rouppe2007,DePontieu2007b}
continuously perturbing the thermodynamic quantities and density
stratification from their equilibrium values. As a result of these
significant deviations from hydrostatic equilibrium, the transition
region is molded by the same dynamic spicular features as the
chromosphere. There is ample evidence that between a height range of
roughly 2,000 and 10,000 km above the photosphere, plasma with
temperatures ranging from 10,000 K to 1 MK co-exist and ``mingle'' in
a highly dynamic fashion. However, the availability of time-averaged,
one-dimensional models of the chromosphere and transition region have
allowed coronal loop models to ignore the complexities and
inherently dynamic nature of this interface \citep[for a review, see][]{Klimchuk2006}. 
Neglecting the intricacies of the chromosphere can be problematic since the chromospheric
heating mechanism requires 50 times more energy than the coronal heating mechanism. 
Consequently, it is perhaps not surprising that a full explanation of coronal heating
remains outstanding, with many unresolved issues and problems \citep[but see][for a different approach]{Aschwanden2007}.

One of these issues that has been problematic for theoretical models of coronal loops is
the large discrepancy between the observed and predicted emission from
the transition region in coronal loops. A variety of theoretical
models predicts emission in transition region lines much in excess of
what is observed with EUV imagers such as SOHO/EIT and TRACE. This
manifests itself in large differences in predicted brightness between
active region loops and transition region moss. Moss is low-lying,
strong EUV emission that was discovered in TRACE EUV images in the Fe~{\sc ix/x} 171 \AA\ and Fe~{\sc xii} 195 \AA\ passbands
\citep{Berger1999,DePontieu1999,Fletcher1999,Berger1999b}. It is the upper
transition region at the base of hot, high pressure loops
\citep{Berger1999,Martens2000}. Observations show
that in a typical Fe~{\sc ix/x} 171 \AA\ EUV image of an active region,
coronal loops and moss regions are roughly of the same brightness.
Despite a variety of approaches, all theoretical models continue to
predict much higher EUV emission from moss regions than from coronal
loops. For example, most coronal loop models based on a steady heating
mechanism predict moss emission that is one to several orders of
magnitude larger than what is observed
\citep{Schrijver2004,Warren2006,Warren2008}. While the
introduction of models based on time-dependent nanoflare heating
\citep[e.g.][]{Patsourakos2008} on average shows a smaller
discrepancy, they still predict moss emission that is 10 times
brighter than what is sometimes observed. Other models remove the discrepancy by introducing a filling factor in the transition region as a free parameter \citep{Winebarger2008}. 

However, none of these models take into account the fact that the transition region emission
in moss regions as observed by EUV imagers such as TRACE and EIT
undergoes significant absorption by neutral gas through Lyman continuum absorption.
Radiation of EUV emission below the head of the hydrogen Lyman continuum at 912 \AA\ can be absorbed by neutral hydrogen \citep{Anzer2005}. In addition, neutral helium can absorb EUV radiation with wavelengths shorter than 504 \AA\, and singly ionized helium can contribute to absorption for wavelengths shorter than 228 \AA. The effects of neutral hydrogen and helium and singly ionized helium on EUV radiation were studied extensively in the past \citep{Schmahl1979, Doschek1982, Mariska1992,Daw1995, Judge1995, Anzer2005} for a variety of solar regions including quiet Sun, coronal holes, prominences, the limb, and active regions, often with confusing results and interpretations. Part of this confusion likely stems from the fact that the amount of neutral plasma in observations of such a wide variety of features is highly variable and most often not directly observed, but simply inferred from the EUV emission. This is not the case for moss: high resolution observations made in the EUV (with TRACE) and chromospheric lines such as H$\alpha$ (with ground-based telescopes) clearly show the presence of neutral plasma at heights where EUV emission occurs \citep{Berger1999}. In fact, moss shows up as a reticulated pattern above plage regions precisely because the EUV emission from the 1-1.5 MK upper
transition region plasma is absorbed by mostly neutral gas in dense,
chromospheric jets or fibrils that occur at similar heights and in the
immediate surroundings of the upper transition region plasma \citep{DePontieu1999}.

The mingling of chromospheric and TR plasma at low heights has been
confirmed by several observational studies in the past few years. For
example, \citet{DePontieu2003} found that in moss regions the
linecenter intensity of one of the dominant chromospheric lines
(H$\alpha$) is very well correlated, both spatially and temporally,
with Fe~{\sc ix/x} EUV emission observed with TRACE. Such a
correlation is difficult to understand unless chromospheric and TR
plasma occur at similar heights. This is confirmed by
\citet{DePontieu2003b} who found that oscillations in moss, observed
in EUV with TRACE, are associated with oscillations in the wings of
H$\alpha$. These correlated oscillations can be explained by
absorption of EUV emission by chromospheric fibrils that
quasi-periodically rise high enough so that they can obscure the
lower-lying upper TR emission from neighboring field lines
\citep{DePontieu2004}. Such a scenario is supported by extensive
observational and modelling work that has shown that these dynamic
fibrils are caused by shock waves that drive chromospheric plasma
upward to reach heights of several 1,000 km
\citep{DePontieu2004,DePontieu2006,Hansteen2006,DePontieu2007,Heggland2007}.

An accurate determination of this absorption in moss regions is thus
clearly an important ingredient in determining the accuracy (or
constraining parameters such as filling factors) of coronal
loop models. In this paper we focus on determining what fraction of
the EUV emission in moss regions is absorbed, so that future work can
directly compare predicted intensities with observed intensities that
have been corrected for this absorption.

We use the density sensitivity of emission lines (of the same ion)
formed above and below the Lyman continuum and compare the obtained
densities from several line pairs. We describe the observations with
Hinode \citep{Kosugi2007} and SOHO \citep{Domingo1995} and the
co-alignment of the EIS \citep{Culhane2007} and SUMER
\citep{Wilhelm1995} spectra in \S 2. In \S 3 we describe the
discrepancies in density measurements for coronal loops and moss
regions, and estimate the amount of absorption of TR emission. We also
observe the same moss region with SECCHI \citep{Howard2008} onboard
the STEREO A and B spacecraft \citep{Kaiser2008} and use the
dependence on viewing angle of the TR emission to estimate the
center-to-limb variation of the absorption (\S 4). We demonstrate that
such absorption compares well with what we detect in numerical
simulations (\S 5). We finish by summarizing our results and briefly
discussing the potential impact of our results on coronal loop
modelling in \S 6.

\section{Details of EIS/SUMER Observations}\label{obs}

To study the differential absorption of EUV emission with wavelengths
below and above 912 \AA, we focus on emission lines from the same ion
that is formed, under ionization equilibrium conditions, at
temperatures that are high enough to clearly show moss ($\log T =
5.7-6.3$). These conditions are quite stringent since most
high-temperature lines above 912 \AA\ are forbidden lines and quite
weak. In addition, the small-scale nature of moss structuring requires
1-2\arcsec~class spectra in both lines. The launch of Hinode and
its EIS instrument has opened up access to a range of Fe~{\sc xii} lines
(186.88 and 195.1 \AA\ in particular) that together with SUMER's Fe~{\sc xii}
1242 \AA\ line provide the tools we require to address the absorption
of EUV moss emission by neutral hydrogen and helium (and singly
ionized helium). We have chosen these lines because the 186.88 \AA\ and
195.1 \AA\ line pair is sensitive to densities in the range of $\log n_e =
8-10.5$, an ideal range for active regions. In addition, the ratio of
195 to 1242 \AA\ is also sensitive to density, so that we can directly
compare the densities derived from both line pairs.

We use an EIS raster of a weak active region (which lacks a NOAA number) taken from 16:44-17:50
UTC on 14-Nov-2007 (see panels A, B of Fig. \ref{f1}). The active region was observed at a heliocentric position of (+150\arcsec, -100\arcsec). The raster
consists of spectra of a range of EUV emission lines (including Fe~{\sc xii}
186.88 and 195.1 \AA) at 256 positions (spaced 1\arcsec~apart in solar
x) observed from west to east at a cadence of 15 s. The data is
despiked (i.e., cosmic ray removal) and calibrated using
{\tt eis\_prep.pro} in the EIS solarsoft software tree.  For each location
in the raster, we calculate the total intensity in the Fe~{\sc xii} lines by
summing over a range of spectral pixels that includes the full line,
and subtracting an estimate of the continuum emission in a spectral
region in the immediate spectral vicinity that is free of emission
lines. The absolute calibration of EIS is described in
\citet{Culhane2007} and is estimated to be accurate to about 20\%.  We
remove the pointing offset between the EIS rasters by using a
cross-correlation technique between the two Fe~{\sc xii} rasters.

During the same time frame SUMER observed this active region in a
range of UV lines that include the Fe~{\sc xii} 1242 \AA\ emission line.
Similarly to both EIS lines, it has a peak in the contribution function at
$\log T=6.15$. The raster was taken from east to west and took much
longer to complete (from 17:12 UTC through 20:35 UTC) because of the
different SUMER sensitivity and the weakness of the Fe~{\sc xii} line. The
raster steps are 1.125\arcsec~with three exposures of 30 s at each
raster position to ensure an adequate count rate. To calibrate the SUMER
data we perform the steps described in \citet{McIntosh2007} which includes
corrections for flatfielding, as well as the absolute calibration. The
last available flat-field is from 2001, which means that there are
some uncertainties in correcting for detector artifacts. This is
clearly visible in panel D of Fig.  \ref{f1} which shows the SUMER Fe~{\sc xii}
1242 \AA\ intensity (calculated in the same way as described for EIS in
the above). The uncertainty in flat-fielding is evidenced by the
presence of bright and dark horizontal streaks in the raster.

To coalign these rasters we perform the following steps. Since these
rasters have very different durations, they are subject to
substantially different solar rotation effects with the features
rastered by SUMER rotating by 30\arcsec, with EIS only undergoing
10\arcsec. To remove this effect, we recalculate the pointing
information for each position of the EIS and SUMER rasters back to the
same reference time $t_0$. In other words, we calculate which features
on the Sun at $t_0$ occupy the SUMER and EIS slits at the diverse
range of positions and times of the rasters. Using this new pointing
information, we then calculate which SUMER raster positions are
nearest in space to the EIS raster positions, and form a SUMER raster
that has the same spatial scale and number of positions as the EIS
rasters. As a final step we perform a final coalignment between the
new SUMER raster and the EIS rasters through visual comparison of the
features in the eastern half of the overlapping parts of the field of
view. We focus on the eastern half because the time difference between
the SUMER and EIS exposures in that area of the field of view are the
smallest, of order $\pm$ 1 hour (see bottom panel of Fig. \ref{f2}),
which increases the likelihood of viewing the same coronal and TR
structures.

The resulting coalignment, shown in Fig. \ref{f1}, is quite
satisfactory, especially in the eastern half of the raster. We use
simultaneous TRACE and XRT data to identify the locations of TR moss.
The high spatial resolution of the TRACE Fe~{\sc ix/x} 171 \AA\ image
(formed at slightly lower temperatures of 1 MK, and taken around 17:45
UTC) shows the reticulated pattern that is typical for TR moss.  This
pattern is caused by the presence of EUV absorbing chromospheric
features; typically jets or dynamic fibrils \citep{DePontieu2003,DePontieu2003b,DePontieu2004,Hansteen2006,
  DePontieu2007}. We find several patches of moss around
$x=20-40$ and $y=0-70$ ($MA$), as well as $x=75-120$ and $y=60-90$ ($MB$). Those
footpoint locations fit very well with the location of the associated
hotter loops in the two XRT images in Fig. \ref{f2}. In addition,
comparison of the Fe~{\sc xii} rasters with the XRT images allows us to
identify coronal loops in the Fe~{\sc xii} rasters around $x=10-40$ and
$y=80-120$ ($A$), as well as $x=60-120$ and $y=30-60$ ($B$), and some smaller
loops around $x=10-25$ and $y=0-15$ ($C$), $x=20$ and $y=25-35$ ($D$),
$x=60-80$ and $y=65-85$ ($E$), and $x=40-65$ and $y=60-75$ ($F$).

The XRT images (Fig. \ref{f2}) are taken at roughly the beginning and
end of the SUMER raster and show the significant changes the western
half of the active region undergoes during the 3 hour raster. We
draw attention to the formation of moss footpoints around $x=80-120$
and $y=60-70$ and associated loops just south of that region. These
changes all occur after the EIS raster has ended. This explains the
different appearance in panels A, B and D of Fig. \ref{f1} of the Fe~{\sc xii}
emission in the southwestern part of the field of view. The eastern part of the active region shows much less variability with stable moss patterns. This is the region we will focus on in the following.


\section{Analysis of EIS/SUMER Observations}\label{anobs}

\begin{figure}
\epsscale{1.0}
\plotone{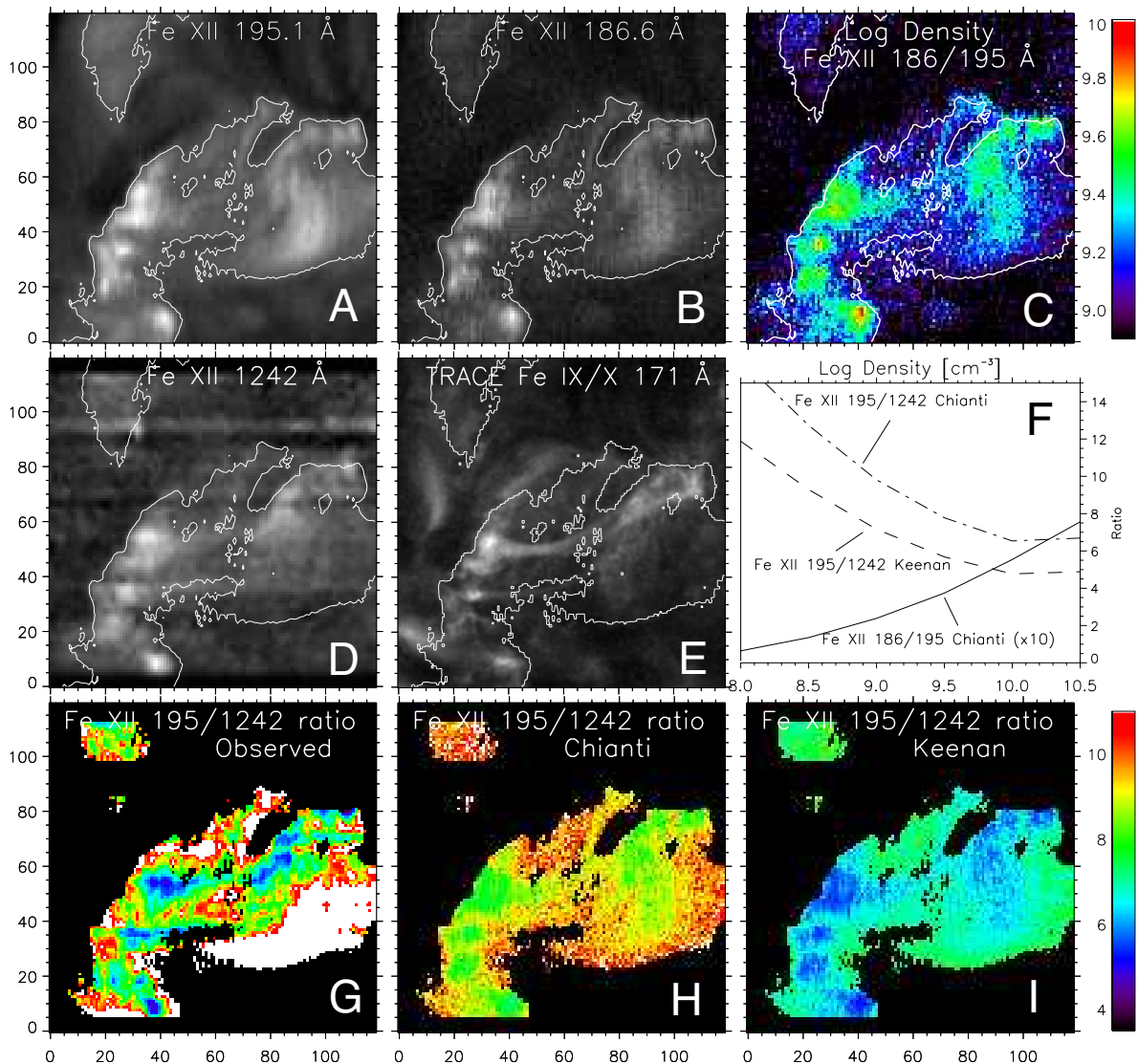}
\caption{Top row shows a spectroheliogram in Fe~{\sc xii} 195.1 \AA\ and
  186.88 \AA\, taken on 14-Nov-2007 from 16:44 to 17:50 UTC with EIS,
  as well as the logarithm of the density as determined from the line
  ratio of the 186.88 and 195.1 \AA\ lines. The middle row shows the a
  slightly spatially smoothed SUMER raster taken from 17:12 through
  20:33 UT and a TRACE Fe~{\sc ix/x} 171 \AA\ image. The middle right panel
  shows the line ratio to density functions we used to determine the
  density for both line pairs. The Fe~{\sc xii} 186/195 \AA\ line ratio is
  multiplied by 10 for illustration purposes. The bottom row shows the
  observed Fe~{\sc xii} 195/1242 \AA\ line ratio, as well as the line ratio that
  is predicted if we use the density determination of the 186/195 line
  ratio, for two atomic models: CHIANTI and that of \citet{Keenan1990}.
  The overplotted contours correspond to a brightness threshold in the
  EIS Fe~{\sc xii} 195.1 \AA raster. See text for details on the
  interpretation.
 \label{f1}}
\end{figure}

\begin{figure}
\epsscale{0.4}
\plotone{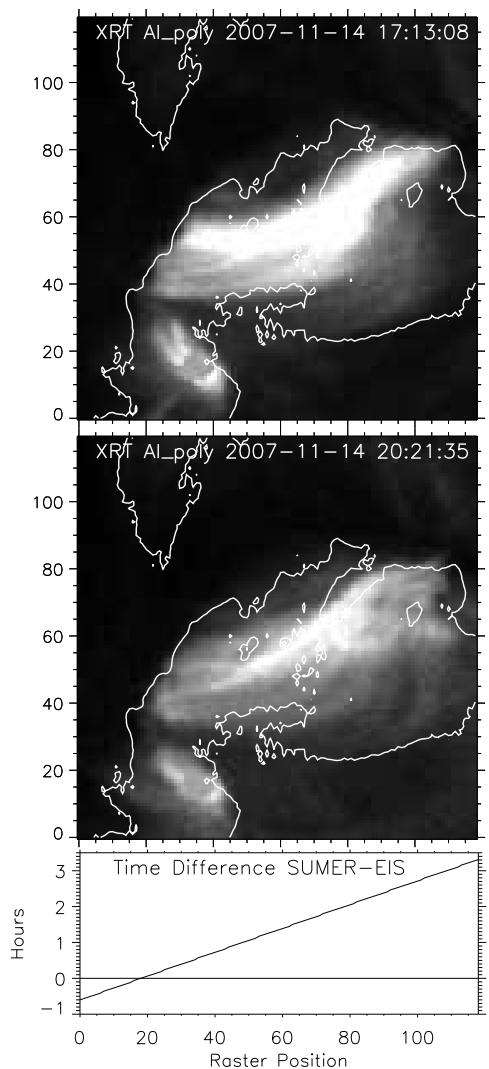}
\caption{Top two images show XRT images ($Al_{poly}$) taken at the
  beginning and end of the SUMER raster. The overplotted contours
  again correspond to the same brightness threshold in the EIS Fe~{\sc xii}
  195 \AA\ raster shown in Figure \ref{f1}. Since the XRT filter is
  sensitive to hotter plasma than the Fe~{\sc xii} rasters, we see much more
  emission from the hot loops connecting the TR moss footpoints. In
  addition, we see that the loop structure in the middle right of the
  image drastically changes in morphology with new moss footpoints
  appearing around $x=80-120$, $y=60$. The lower panel shows, in
  hours, the time difference between the SUMER and EIS exposures
  (shown in Fig. \ref{f1}) at each coaligned raster location. The time
  differences are caused by the different durations of the SUMER and
  EIS rasters, and the fact that they raster in E-W and W-E directions
  respectively. The left part of the rasters is taken within 1 hour of
  each other, whereas the right part ($x>70$) shows exposures that are
  taken 2 or more hours apart. See text for details.\label{f2}}
\end{figure}

We calculate the density sensitivity of the Fe~{\sc xii} 186.88 and 195.1
\AA\ lines by using Chianti \citep{Dere1997}. The line ratio is
sensitive to densities between $\log \, n_e$=8.5 and 10.5, as shown in
panel F of Fig. \ref{f1}. We calculate the intensities as described in \S \ref{obs}, 
and then determine the ratio of the 186.88 \AA\ and
195.1 \AA\ intensity. We use the curve in panel F to calculate the density
shown in panel C of Fig. \ref{f1}. We find densities of order $10^9$
cm$^{-3}$ in the coronal loops we have identified earlier, with
significantly higher densities of order $4 \, 10^9$- $10^{10}$
cm$^{-3}$ in the moss regions. These values are very reasonable, and
similar to those found in active region loops and moss regions
reported by \citet{Fletcher1999,Warren2008} using CDS and EIS
respectively. 

We can use the densities determined from this line ratio to
calculate what the intensity ratio between the 195 \AA\ and 1242 \AA\ lines
should be, given the density sensitivity of the latter ratio. This
density sensitivity is not as straightforward to use as the 186/195
ratio. Firstly, there is some discussion regarding the atomic physics
underlying the density sensitivity, with \citet{Keenan1990} and CHIANTI
\citep{Dere1997} giving slightly different curves of line ratio to
density (see panel F of Fig. \ref{f1}), with the \citet{Keenan1990} results showing lower
ratios by a factor of 0.73. In what follows we will use both and
compare the results. Using the density derived from 186/195 \AA\ and the
CHIANTI and \citet{Keenan1990} curves, we now calculate for each position in the
rasters the predicted Fe~{\sc xii} 195/1242 \AA\ intensity ratio (shown in panels H
and I of Fig. \ref{f1}). The values of this ratio vary between 8 and
12 for the CHIANTI curve, and 5 and 8 for the \citet{Keenan1990} curve. 

We now calculate the Fe~{\sc xii} 195/1242 \AA\ ratio directly from the
observations using the coaligned rasters. The Fe~{\sc xii} 1242 \AA\ raster is
quite noisy because of the low count rates, so we use a boxcar
smoothing over 3 pixels to remove some of that noise when calculating
the ratio between the EIS and SUMER rasters. The resulting ratio is
shown in panel G of Fig. \ref{f1} with the same color table as panels
H and I. The observed ratio varies between 5 and 15 and generally shows
a much better correspondence with the levels predicted by the CHIANTI
curve. However, it is immediately clear that there are also
significant differences between the predicted and observed intensity
ratios. The most striking difference is that the moss regions in the
eastern half of the field of view systematically show lower ratios (of
order 4 to 6) than predicted (of order 8-10). This implies that the EUV emission with wavelengths below 912 \AA\ is weaker by a factor of order 2 in moss regions than would be expected. In contrast the loop
structures we identified in the above show a much better
correspondence with typically values around 10 (yellow) in both the
observed and predicted maps, especially for loops $A$, $C$, $D$ and
$E$. 

\begin{figure}
\epsscale{0.6}
\plotone{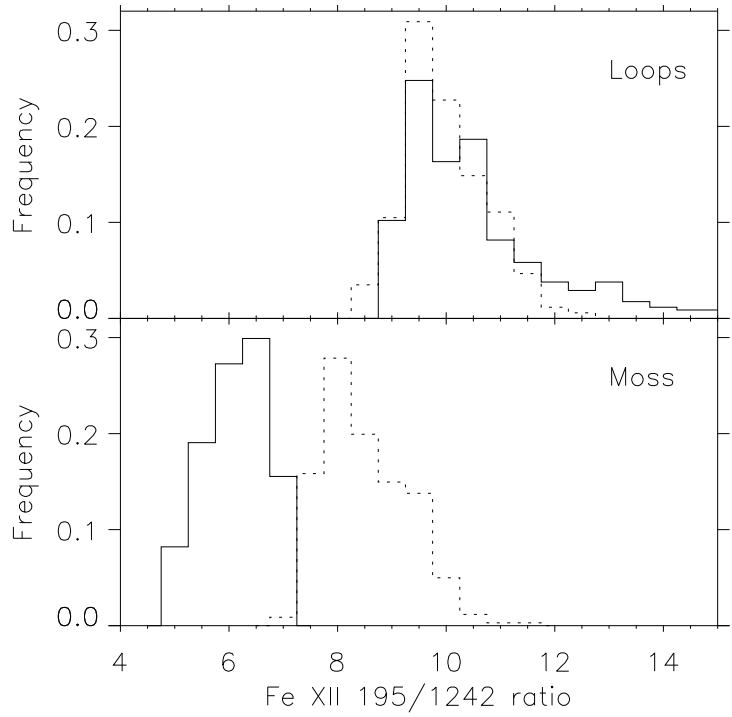}
\caption{Comparison of histograms of observed (full lines) and predicted (dotted lines) values of the 195.1 \AA\ to 1242 \AA\ intensity ratio in coronal loops $A$ and $F$ (top row), and moss region $MA$ (bottom row). See \S~\ref{obs} for the definition and location of loop and moss regions.
 \label{f1b}}
\end{figure}

Fig.~\ref{f1b} shows these differences in more detail by comparing histograms of the predicted and observed 195/1242 intensity ratio for coronal loops $A$ and $F$ (top panel), as well as moss region $MA$ (bottom panel). As we can see, the loop regions show reasonable correspondence between the predicted and observed ratios, whereas the moss regions show a significant difference of order 2. 

The loop regions often show higher observed values than predicted. This can be seen in Fig.~\ref{f1b}, but also in the large loops $B$ in Fig.~\ref{f1}.
We believe this is likely because the active region thermal structure
in the western half of the field of view changed towards the end of
the SUMER raster, so that the Fe~{\sc xii} 1242 \AA\ emission becomes much
lower, perhaps because of a heating of the loops in that region
towards the end of the SUMER raster (which is suggested by the XRT
timeseries, Fig. \ref{f2}). In this scenario, the EIS raster, taken 3
hours earlier then still shows evidence of the cooler loops.

Are the differences between the observed and predicted ratios caused
by Lyman continuum absorption in the moss regions of 195 \AA\ emission? 
Our results hinge on the fact that the observed spatial variation of the 195/1242 \AA\ ratio between moss and loops is larger than predicted by densities determined from the 186/195 \AA\ ratio. We observe a generally much higher contrast and lower values in the map of observed ratios. Several mechanisms could
potentially contribute to this mismatch between predicted and observed ratios. We argue 
in what follows that none of these can explain our observations and that Lyman continuum absorption 
is the most likely cause.

\begin{figure*}
\epsscale{0.5}
\plotone{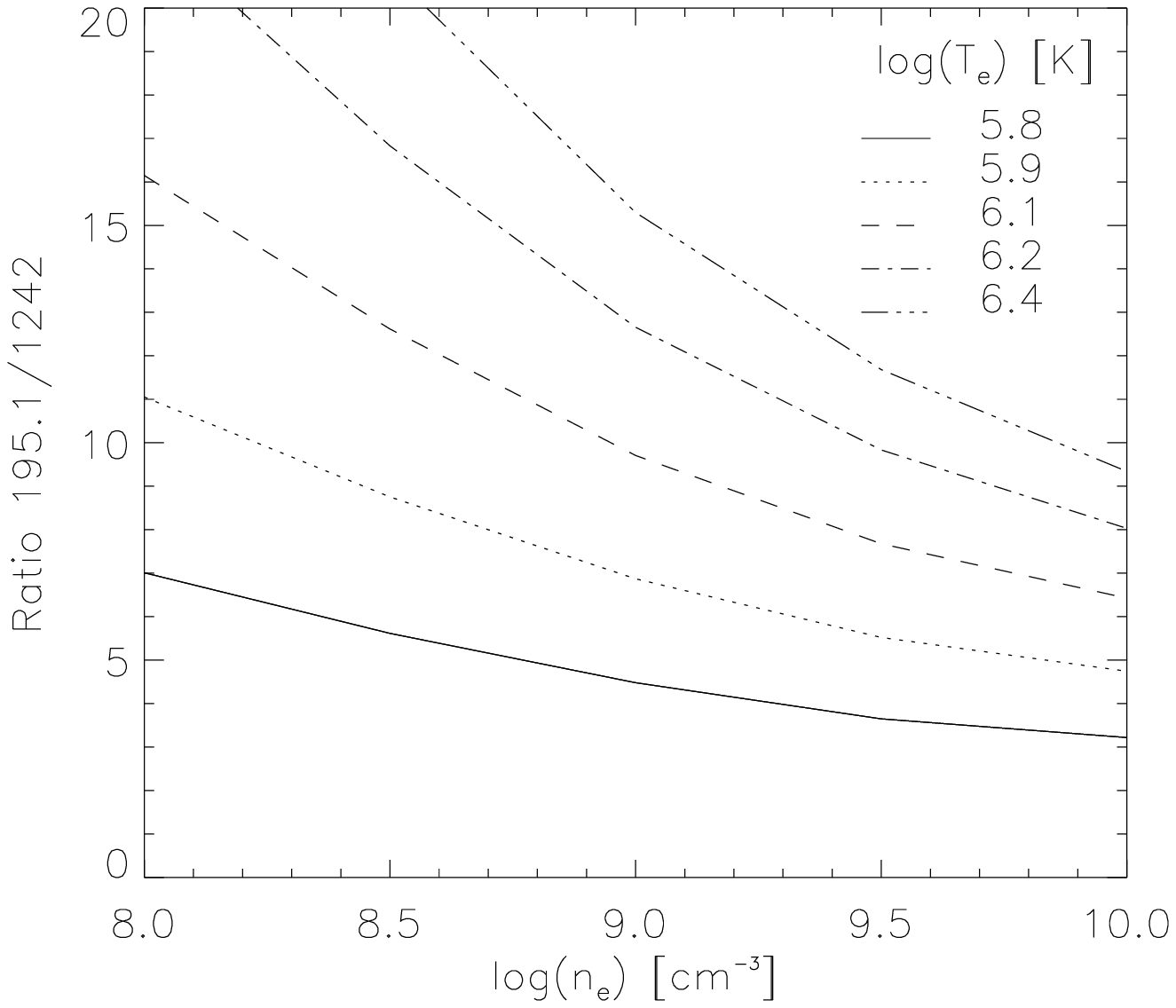}
\plotone{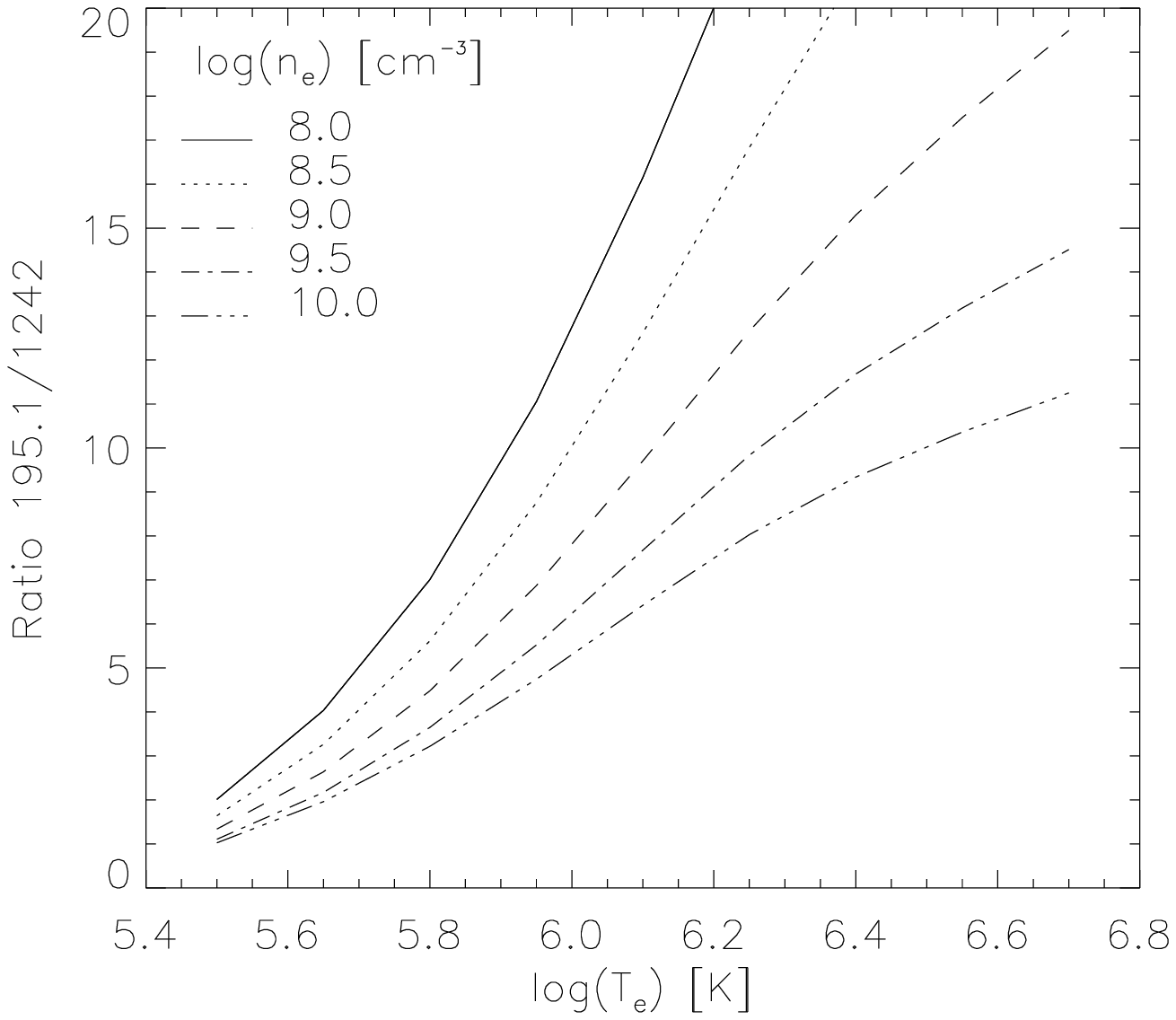}
\caption{The density sensitive intensity ratio of the Fe~{\sc xii} 195 \AA\
  and 1242 \AA\ lines also varies depending on the temperature of the
  plasma (form the CHIANTI atomic database). The left plot shows the
  density sensitivity for a range of temperatures with $\log T$ from
  5.8 through 6.4. The temperature associated with the maximum
  formation is $\log T = 6.15$, with the contribution function dropping
  off steeply towards lower and higher temperatures. The right plot
  shows the same dependency, but now the ratio is plotted as a
  function of temperature for a range of densities. The densities in
  the active region studied vary between $\log n_e = 9 - 10$.
\label{f3}}
\end{figure*}

A first effect is the temperature dependence of the 195/1242 \AA\ line ratio. 
This temperature sensitivity was briefly discussed by
\citet{DelZanna2005} and is illustrated in Fig. \ref{f3}. We see from the
top panel that for the range of densities we observe in this active
region ($\log n_e = 9-10$) the 195/1242 \AA\ line ratio can vary from 5 to 3 for
$\log T=5.8$ to from 15 to 10 for $\log T=6.4$. Both of these
temperature values are on the extreme end of what we are most likely
observing in these passbands, since the contribution functions of both
lines are sharply peaked around $\log T=6.15$. 

We believe that in the coronal loops this temperature dependence can explain some of the 
spatial variations of the 195/1242 \AA\ line ratio. In loops, we observe these variations to be larger than predicted. However, we argue that in the moss regions, the temperature dependence does not impact our result.
We know from the 186/195 \AA\ ratio that
the densities in the moss are of order $\log \, n_e = 9.5-10$. In
principle our low observed values of the 195/1242 \AA\ ratio in the moss could thus be
caused by a much lower temperature ($\log \, T=5.8$), since Fig. \ref{f3}
shows ratios of order 3 to 4. However, we know that moss is always
associated with high pressure, hot and dense loops. So, if the upper transition region moss
really had a temperature that is much lower than the peak of the Fe~{\sc xii} contribution function, then the high pressure, dense coronal loops that are associated with the upper TR moss would include plasma with temperatures 
at $\log T=6.15$. Since the loops associated with moss are high pressure and dense and at the temperature
corresponding to the peak of the Fe~{\sc xii} contribution function, these loops would
be very bright in the Fe~{\sc xii} rasters. However, we do not observe any
such loops in or around the mossy regions. Our observations thus indicate that the temperature of
the moss is not low enough to cause the low levels of 195/1242 \AA\
ratios that we observe in the moss. The lack of Fe XII loops connecting to moss has also been observed for other active regions by \citet{Antiochos2003}

Another effect that could contribute to the mismatch between observed and predicted 195/1242 \AA\ ratio in moss regions is the presence of blends. Any effects of blends can only impact our results and conclusion if the blends preferentially occur in loops or moss. We now show that the effect of blends does not change our results. \citet{Young2009} have studied the density sensitivity of the 186.88 \AA\ and 195 \AA\ lines in detail. They find that there are several uncertainties in the determination of the intensities of the Fe~{\sc xii} 186 and 195 \AA\ lines that can potentially impact our analysis. There is a weak line at 186.98 \AA\, just redward of the Fe~{\sc xii} line, which is likely a Ni XI transition \citep{Young2009}. In our analysis, we avoid contributions of this line by excluding the wavelength range around the Ni XI line in our calculation of the total intensity of the Fe XII 186.88 \AA\ line. The Fe~{\sc xii} 186.88 \AA\ is also impacted by a weak blend of S XI at 186.84 \AA\ \citep{Young2009}. Under equilibrium conditions, this line is formed at $\log T$=6.3, slightly higher than the $\log T$ = 6.15 temperature of the Fe~{\sc xii} line. \citet{Young2009} find that the S XI intensity is typically 2-5\% of the Fe~{\sc xii} 186.88 \AA\ line, which is smaller than the photon noise on the observed Fe~{\sc xii} 186.88 \AA\ intensities. Given the density sensitivity of the 186/195 \AA\ ratio (Fig. \ref{f1}), such a small difference cannot lead to a significant reduction in the discrepancy between predicted and observed 195/1242 \AA\ intensity ratio in the moss regions. To remove the factor of 2 discrepancy in 195/1242 \AA\ ratio would require a change in the 195/186 ratio by a similar factor. Clearly the S XI blend is too weak to impact our results. In addition, even if the blend was much stronger than previously observed, it would more likely impact the hotter moss regions, because the S XI line is formed at slighly higher temperatures. That would mean that in the moss the Fe~{\sc xii} 186.88 \AA\ intensity is overestimated by our analysis, which would lead to a predicted 195/1242 \AA\ intensity ratio with even less difference between moss and loops than is observed. 

\citet{Young2009} shows that the Fe~{\sc xii} 195.12 \AA\ line similarly has some weak lines in its immediate vicinity. We again avoid contributions of thes lines by excluding the wavelength range of these weaker lines in our calculation of the total intensity of the Fe~{\sc xii} line. This line is also blended by another Fe~{\sc xii} line, which is at levels of less than 10\% for the densities we find here ($< 10^{10}$ cm$^{-3}$). Given the weakness of the blend, and the fact that the brightness of the blend should not show any difference between moss and loops (since it is also Fe~{\sc xii}), this blend cannot change our conclusions.

\citet{Young2009} have compared densities from different line pairs and found some discrepancies. They suggested that the density determination from the Fe~{\sc xii} 186.88 and 195 \AA\ pair may lead to overestimates of the density by 0.2-0.4 dex, especially in higher density regions. Since our observations show that the moss regions have the highest densities, this currently unexplained effect (most likely due to uncertainties in the atomic data) could
lead us to overestimate the moss densities. However, overestimates of the density in moss would lead to even higher values for the predicted ratio of  Fe~{\sc xii} 195 and 1242 \AA\ intensity. In summary, this issue would not remove, but rather exacerbate the discrepancy between observed and predicted ratios.

We have also investigated the effects of the main sources of noise on the EIS spectra. We calculated the errors on our intensities using the eis\_prep software that is part of the EIS solarsoft tree, and found that photon (Poisson) noise dominates the relative errors. In the moss and loop regions we study, the spectra are high signal to noise, of order 20 for the 186.88 \AA\ line and 30-40 for the 195 \AA\ line. This means that the relative errors $\sigma$ on the intensities are small, of order 5 \%. If we assume a worst case scenario in which, e.g., the 186.88 \AA\ intensity is lowered by $\sigma_{186}$ and the 195 \AA\ intensity is increased by $\sigma_{195}$, the predicted 195/1242 ratio map is shifted to higher values. However, such a shift occurs in equal measure for moss and loop regions, so that the predicted spatial variation of the 195/1242 \AA\ ratio cannot be reconciled with the observed ratio. In other words, while these errors can introduce noise on the predicted ratio, they cannot explain the consistently lower values of the 195/1242 ratio that we observe in the moss regions. A similar argument can be made for the absolute calibration errors between SUMER and EIS. Such errors will not depend on the nature of solar region observed (especially since those are of similar brightness), and cannot explain the spatial variation we observe.

We conclude that the most likely explanation for our EIS/SUMER observations is that EUV emission (at 195 \AA) in upper TR moss is reduced in brightness by a factor of order 2 because of absorption from neutral hydrogen and helium, and singly ionized helium in the chromospheric jets that permeate the moss.

\section{Analysis of STEREO Observations}\label{stereo}

The region we study with EIS and SUMER was observed by Hinode and SOHO close to disk center. However, it was also observed by both STEREO spacecraft (SC). The multiple vantage points these spacecraft provide allow us to also investigate the effect of the angle $\theta$ between the line-of-sight vector and the local vertical on the chromospheric absorption of TR EUV emission. We use simultaneous STEREO A and B images taken in the Fe~{\sc xii} 195 \AA\ passband of SECCHI on 14-Nov-2007 during the EIS and SUMER rasters (i.e., between 16 and 18 UTC). The images were corrected for dark current using the software provided in the STEREO solarsoft package. To account for the different distance to the Sun of both spacecraft, we correct the intensities observed by STEREO A by multiplying the SC A intensities with a factor of $(Rsun_A/Rsun_B)^2=(994/920)^2$, where $Rsun_A$ and $Rsun_B$ is the solar radius in arcsec as seen from SC A \& B respectively.

\begin{figure*}
\epsscale{0.5}
\plotone{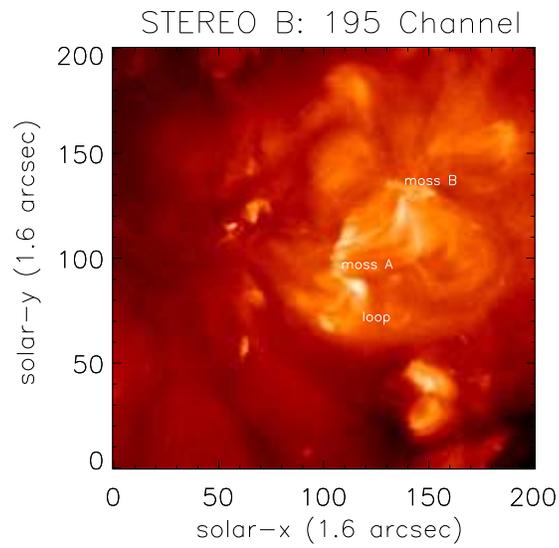}
\caption{Fe~{\sc xii} \AA\ image of the same active region studied with EIS and SUMER, as seen with STEREO B. Marked are three regions of interest (one loop region, and two moss regions) for which the intensities seen with STEREO A and B are compared (see Fig. \ref{f4b}). STEREO A observed this region at disk center, whereas STEREO B observed the region at a viewing angle of 40 degrees (with the local vertical).
\label{f4a}}
\end{figure*}

\begin{figure*}
\epsscale{0.7}
\plotone{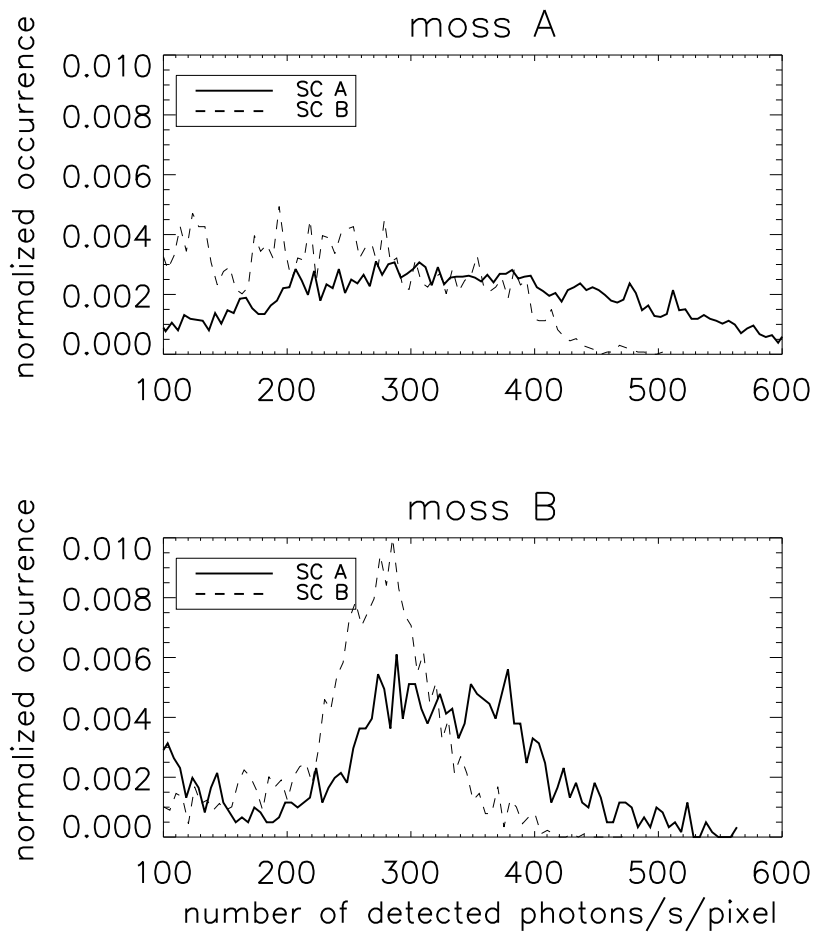}
\caption{Histograms of the intensity in the two moss regions (marked as A and B in Fig. \ref{f4a}) for Fe~{\sc xii} 195 \AA\ observed with STEREO A (full line) and STEREO B (dashed line). Both regions are reduced in intensity by a factor of roughly 0.75 when viewed with STEREO B. This reduction in intensity is caused by Lyman continuum absorption of EUV emission by chromospheric plasma that occurs at similar heights as the EUV emission. Such absorption is increased when the region is viewed from the side, since the observable surface area of chromospheric plasma is increased. 
\label{f4b}}
\end{figure*}

We study the brightness of three regions of interest in this active region as observed with SC A and B. Fig. \ref{f4a} shows a SC B image of the two moss regions we selected as well as a loop-like structure. All of these regions also occur in the EIS and SUMER rasters of Fig. \ref{f1}. The region is seen at disk center by STEREO A, and at an angle of 40 degrees between line-of-sight and local vertical by STEREO B. We determine a histogram of intensities seen by SC A and by SC B for the three regions of interest. We find that the average intensity of the loop is the same for SC A and B (to within 5\%). The observed discrepancy is within the range expected from the absolute calibration error, which is estimated to be of order less than 10\% \footnote{Wuelser, 2008: private communication}. In contrast, the histogram of intensities for both moss regions (Fig. \ref{f4b}) shows significant differences, with the emission seen by SC B reduced to 75\% of that seen by SC A. Moss region A shows an average intensity of 340 (normalized units) with a standard deviation of 130 when seen by SC A, whereas it is weaker at 240$\pm 93$ for SC B. Similarly, moss region B is at $310 \pm 100$ for SC A, and $260 \pm 60$ for SC B. This is confirmed by the histogram plots, which show a marked increase of very bright features seen with SC A (Fig. \ref{f4b}). 

We find that the coronal loop intensities do not change, whereas the moss region intensities decrease significantly when viewed from the side (at a viewing angle of 40 degrees). These results indicate that there is a significant increase of absorption by neutral hydrogen or helium in moss regions when those are viewed from the side. The fact that the loop intensities do not change significantly is to be expected since such loops occur high in the corona, where there is little neutral hydrogen or helium. In contrast, the mix of cool jets and hot plasma in moss regions implies that an observer who views the moss from the side (at a non-negligible viewing angle from the local vertical) will see a larger surface area of chromospheric plasma, which will lead to increased absorption. The amount of absorption seems to roughly scale, perhaps coincidentially, as $\cos \theta$, with $\theta=40$ degrees for our observations.

\section{Analysis of Simulations}\label{sim}

The presence of significant chromospheric absorption of TR EUV emission with wavelengths shorter
than 912 \AA\ is confirmed by advanced radiative MHD simulations. In the Oslo Stagger Code 
\citep{Hansteenetal2007} models of the solar atmosphere stretching from the convection zone to 
corona are constructed. These models include convection, non-grey, non-LTE radiative losses in the 
photosphere, chromosphere, and corona, as well as conduction along the magnetic field. 

An initially potential magnetic field is inserted into a fully convective model of dimensions 
$16\times 8\times 15.5$~Mm$^3$ of which $1.5$~Mm is below the photosphere and $14$~Mm above. The grid size in the horizontal direction is 65 km, with a smaller grid size of 32 km in the vertical direction.
Convective motions and photospheric granular dynamics stress the magnetic field resulting in an 
upwardly propagating Poynting flux that is dissipated in the outer solar layers. After approximately 
20~minutes solar time, and depending on the initial magnetic field strength and topology, 
the dissipated energy is sufficient to raise and maintain the temperature in the upper layers to 
coronal values. Thus a simulated atmospheric structure emerges, with photosphere, chromosphere, 
and corona much as one believes exists on the sun. In the present context we point out that the
chromosphere --- corona interface, the transition region, in this model is extremely corrugated 
ranging in height from some 1.5~Mm to $5-6$~Mm above the photosphere. Thus at any given time we
find plasma at temperatures ranging from $5,000$~K to $1$~MK in the same height range. As a 
consequence cool plasma in sufficient quantities to absorb short wavelength EUV radiation is found
at these heights.

To quantify this we calculate the intensity in both the Fe~{\sc xii} 1242 \AA\ and 195 \AA\ lines from a 
given snapshot. This is accomplished by integrating the optically thin contribution function 
\[
dI=A_{\rm el}n_{\rm e}^2g(T)e^{-\tau}dz
\]
along a rays parallel to the ``line of sight'', where $A_{\rm el}$ is the element abundance relative 
hydrogen, $n_{\rm e}$ is the electron density, $\tau$ is the optical depth of the absorbing opacity, 
and $g(T)$ is a function that incorporates the ionization state of the radiating material and the 
collisional excitation rates:
\[
g(T)=0.83 h\nu n_{j}A_{ji}f(T)/n_{\rm e}.
\]
Here $h\nu$ is the energy of the transition, $n_{j}$ is the population of the upper level of the 
radiating ion, $A_{ji}$ the Einstein coefficient for the transition, and $f(T)$ the ionization state 
of the radiating ion \citep{Dere1997,Landi2006}. The optical depth $\tau$ is 
computed assuming that all absorption comes from neutral hydrogen and neutral and singly ionized 
helium such that following \citet{Anzer2005} we write
\[ 
\tau=\sigma_{\rm H~I} N_{\rm H~I}+\sigma_{\rm He~I}N_{\rm He~II}
+\sigma_{\rm He~II}N_{\rm He~II}
\]
where $N_{\rm H~I}$ is the total column density of neutral hydrogen along the line of sight, and 
$N_{\rm He~I}$ and $N_{\rm He~II}$ are those of neutral and singly ionized helium. The 
$\sigma$-values are the respective photoinization cross sections which depend only on wavelength. In 
computing the column densities of hydrogen and helium we assume the ionization derived by 
\citet{Mazzotta1998} as found in the CHIANTI package. 

\begin{figure}
\epsscale{0.7}
\plotone{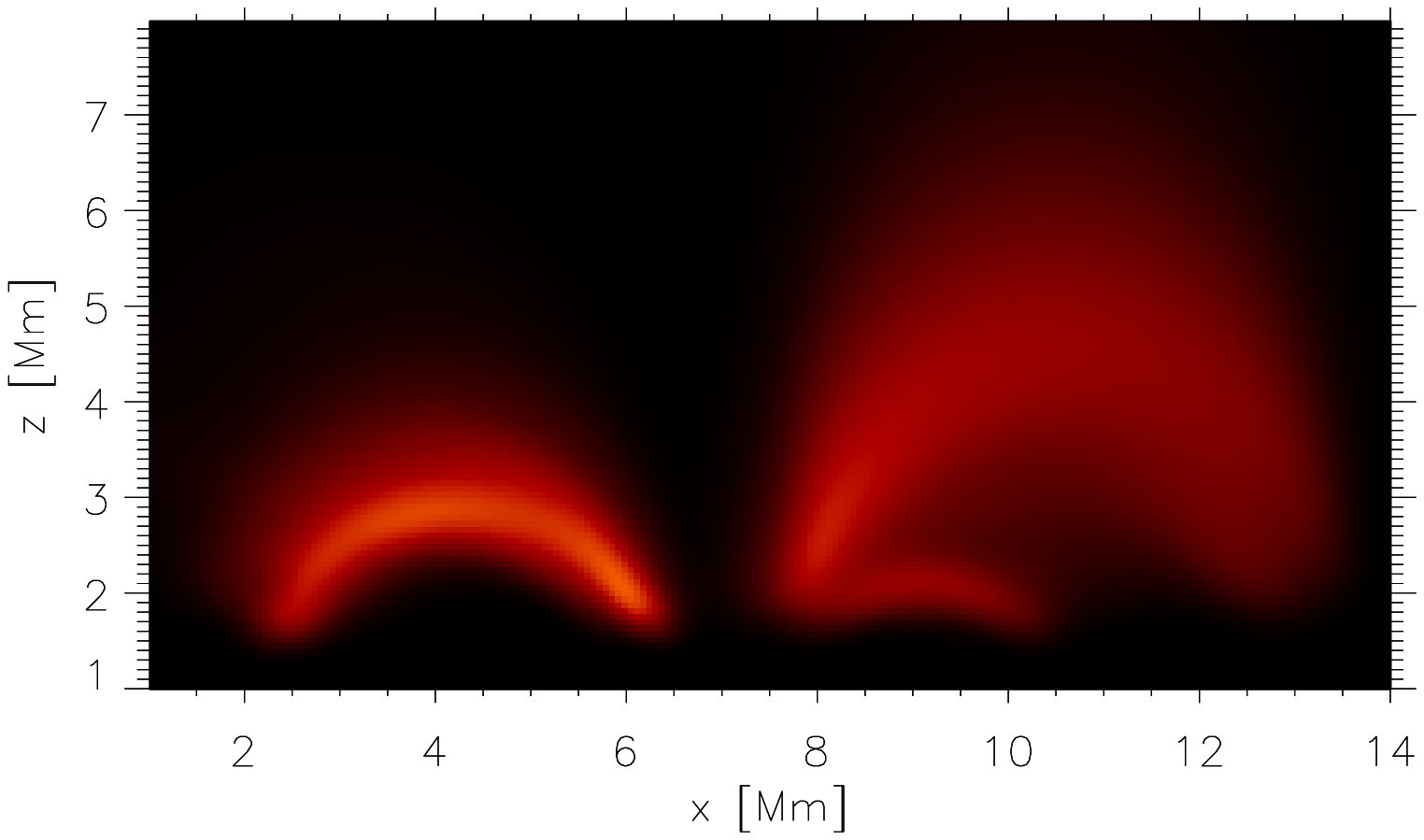}
\plotone{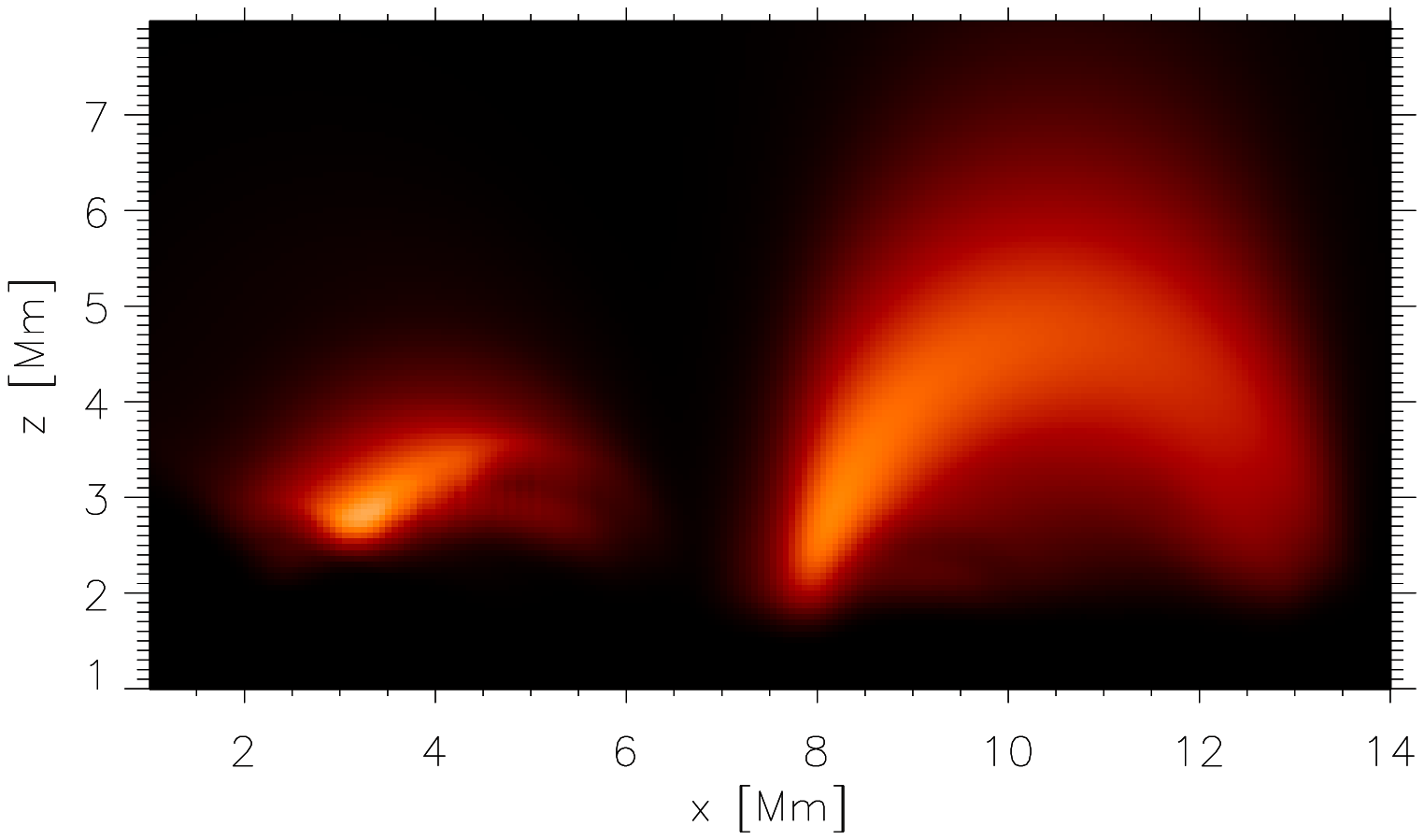}
\caption{Side views of synthetic Fe~{\sc xii} intensities from our 3D
  radiative MHD simulation. The intensity shown in each panel is
  summed along a ray parallel to the y-axis (perpendicular to the
  plane of the figure) using the density and temperature in the
  simulated box, and CHIANTI to calculate the emergent intensity. The
  top panel shows Fe~{\sc xii} 1242 \AA\ emission, whereas the bottom panel
  shows Fe~{\sc xii} 195 \AA\ emission. Notice the significant differences
  between the intensities, both in the higher-lying loops (because of
  density sensitivity), and at the lower regions or footpoints of the
  loops ($z < 3$ Mm) because of the absorption of EUV photons that
  have wavelengths below 912 \AA. See text for details. \label{f5}}
\end{figure}

In figure~\ref{f5} we plot the emergent intensities of the Fe~{\sc xii} 1242~{\AA} and
195~{\AA} lines as seen from the side, {\it e.g.} as seen at the limb. 
Notice the significant differences between the intensities, both in the higher-lying loops 
(because of density sensitivity), and at the lower regions near the footpoints of the loops (z $<$
3 Mm) because of the chromospheric absorption of EUV photons that have wavelengths below 912 \AA.
The presence of almost complete extinction of the 195 \AA\ intensity in the lower 1-3 Mm of the coronal loops is caused by absorption from upper chromospheric extrusions into coronal territory.

\section{Summary}\label{discuss}

We find evidence for significant absorption of EUV emission (with wavelengths below 912 \AA) in the transition region of coronal loops because of the presence at similar heights of dense, chromospheric gas. We use the Fe~{\sc xii} 186.88, 195 and 1242 \AA\ lines from quasi-simultaneous Hinode/EIS and SOHO/SUMER spectra of an active region and show that the transition region emission in moss regions is reduced by a factor of order 2 because of absorption by neutral hydrogen and helium and singly ionized helium that occurs at similar heights in chromospheric jets. We use STEREO/SECCHI observations of the same active region using both spacecraft and find that this absorption is further increased when moss regions are viewed closer to the limb. These observational results are confirmed by advanced 3D numerical simulations which show similar levels of absorption because of co-mingling of cool, chromospheric plasma and hot, upper transition region plasma at similar heights.

The determination of this type of absorption significantly reduces or removes an uncertainty that has plagued comparisons of EUV images with numerical models of coronal loops, which have had trouble reconciling the observed and predicted EUV brightness at the footpoints of loops. Current models typically show a mismatch of one order of magnitude. A variety of attempts at reducing the discrepancy include height (and magnetic field) dependent filling
factors, time-dependent heating models, etc... By determining what the real, unobstructed emission in the upper transition region of coronal loops is, our results reduce (but do not totally remove) the current discrepancy and will guide modelers towards better constraints on their free parameters and assumptions. The STEREO observations and numerical results suggest that full Sun models need to also consider the center-to-limb variation of the absorption.

The significant absorption of EUV emission in the transition region by cool plasma in chromospheric jets indicates that coronal loop modellers should be aware of and take into account the properties of the absorbing jets. For example, is it possible that the observed inverse correlation between ``filling factor'' and coronal loop pressure reported by \citet{Warren2008} is related to the dependence on coronal pressure of the height range over which the transition region occurs in combination with the presence of absorbing jets? Given the significant absorption by chromospheric jets, more attention is also warranted for the effects of chromospheric absorption on the detailed shape of line profiles of coronal emission lines \citep[see, e.g.][]{Spadaro1996}.

The total amount of absorption we find is of order at least 2 to 3 (for viewing angles of less than 40 degrees from disk center). Further observations with STEREO/SECCHI for a wider variety of viewing angles have the potential of revealing the true center to limb variation of moss brightness, which can be compared to that of numerical simulations, and may reveal details about the filling factor and orientation of chromospheric features in the moss phenomenon. In addition, simultaneous observations with EIS and SUMER of a larger statistical sample of active regions will help determine the variability of this absorption across active regions of varying activity.

\acknowledgements
SWM acknowledges support from NSF grant ATM-0541567 and NASA grants
NNG06GC89G and NNX08AH45G. BDP is supported by NASA grants NAS5-38099 (TRACE),
NNM07AA01C (HINODE), NNG06GG79G and NNX08AH45G. We would like to thank Amy Winebarger and Peter Young for helpful discussions. {\em Hinode} is a Japanese mission
developed and launched by ISAS/JAXA, with NAOJ as a domestic partner
and NASA and STFC (UK) as international partners. The SECCHI data were produced 
by an international consortium of the Naval Research Laboratory (USA), Lockheed Martin Solar and
Astrophysics Lab (USA), NASA Goddard Space Flight Center (USA),
Rutherford Appleton Laboratory (UK), University of Birmingham (UK),
Max$-$Planck$-$Institut for Solar System Research (Germany), Centre
Spatiale de Li\`ege (Belgium), Institut d'Optique Th\'eorique et
Applique\'e (France), and Institut d'Astrophysique Spatiale (France).

\bibliographystyle{apj}


\begin{thebibliography}{44}
\expandafter\ifx\csname natexlab\endcsname\relax\def\natexlab#1{#1}\fi

\bibitem[{{Antiochos} {et~al.}(2003){Antiochos}, {Karpen}, {DeLuca}, {Golub},
  \& {Hamilton}}]{Antiochos2003}
{Antiochos}, S.~K., {Karpen}, J.~T., {DeLuca}, E.~E., {Golub}, L., \&
  {Hamilton}, P. 2003, \apj, 590, 547

\bibitem[{{Anzer} \& {Heinzel}(2005)}]{Anzer2005}
{Anzer}, U. \& {Heinzel}, P. 2005, \apj, 622, 714

\bibitem[{{Aschwanden} {et~al.}(2007){Aschwanden}, {Winebarger}, {Tsiklauri},
  \& {Peter}}]{Aschwanden2007}
{Aschwanden}, M.~J., {Winebarger}, A., {Tsiklauri}, D., \& {Peter}, H. 2007,
  \apj, 659, 1673

\bibitem[{{Berger} {et~al.}(1999{\natexlab{a}}){Berger}, {De Pontieu},
  {Fletcher}, {Schrijver}, {Tarbell}, \& {Title}}]{Berger1999b}
{Berger}, T.~E., {De Pontieu}, B., {Fletcher}, L., {Schrijver}, C.~J.,
  {Tarbell}, T.~D., \& {Title}, A.~M. 1999{\natexlab{a}}, \solphys, 190, 409

\bibitem[{{Berger} {et~al.}(1999{\natexlab{b}}){Berger}, {De Pontieu},
  {Schrijver}, \& {Title}}]{Berger1999}
{Berger}, T.~E., {De Pontieu}, B., {Schrijver}, C.~J., \& {Title}, A.~M.
  1999{\natexlab{b}}, \apjl, 519, L97

\bibitem[{{Culhane} {et~al.}(2007){Culhane}, {Harra}, {James}, {Al-Janabi},
  {Bradley}, {Chaudry}, {Rees}, {Tandy}, {Thomas}, {Whillock}, {Winter},
  {Doschek}, {Korendyke}, {Brown}, {Myers}, {Mariska}, {Seely}, {Lang}, {Kent},
  {Shaughnessy}, {Young}, {Simnett}, {Castelli}, {Mahmoud}, {Mapson-Menard},
  {Probyn}, {Thomas}, {Davila}, {Dere}, {Windt}, {Shea}, {Hagood}, {Moye},
  {Hara}, {Watanabe}, {Matsuzaki}, {Kosugi}, {Hansteen}, \&
  {Wikstol}}]{Culhane2007}
{Culhane}, J.~L., {Harra}, L.~K., {James}, A.~M., {Al-Janabi}, K., {Bradley},
  L.~J., {Chaudry}, R.~A., {Rees}, K., {Tandy}, J.~A., {Thomas}, P.,
  {Whillock}, M.~C.~R., {Winter}, B., {Doschek}, G.~A., {Korendyke}, C.~M.,
  {Brown}, C.~M., {Myers}, S., {Mariska}, J., {Seely}, J., {Lang}, J., {Kent},
  B.~J., {Shaughnessy}, B.~M., {Young}, P.~R., {Simnett}, G.~M., {Castelli},
  C.~M., {Mahmoud}, S., {Mapson-Menard}, H., {Probyn}, B.~J., {Thomas}, R.~J.,
  {Davila}, J., {Dere}, K., {Windt}, D., {Shea}, J., {Hagood}, R., {Moye}, R.,
  {Hara}, H., {Watanabe}, T., {Matsuzaki}, K., {Kosugi}, T., {Hansteen}, V., \&
  {Wikstol}, {\O}. 2007, \solphys, 243, 19

\bibitem[{{Daw} {et~al.}(1995){Daw}, {Deluca}, \& {Golub}}]{Daw1995}
{Daw}, A., {Deluca}, E.~E., \& {Golub}, L. 1995, \apj, 453, 929

\bibitem[{{De Pontieu} {et~al.}(1999){De Pontieu}, {Berger}, {Schrijver}, \&
  {Title}}]{DePontieu1999}
{De Pontieu}, B., {Berger}, T.~E., {Schrijver}, C.~J., \& {Title}, A.~M. 1999,
  \solphys, 190, 419

\bibitem[{{De Pontieu} \& {Erd{\'e}lyi}(2006)}]{DePontieu2006}
{De Pontieu}, B. \& {Erd{\'e}lyi}, R. 2006, Royal Society of London
  Philosophical Transactions Series A, 364, 383

\bibitem[{{De Pontieu} {et~al.}(2003{\natexlab{a}}){De Pontieu}, {Erd{\'e}lyi},
  \& {de Wijn}}]{DePontieu2003b}
{De Pontieu}, B., {Erd{\'e}lyi}, R., \& {de Wijn}, A.~G. 2003{\natexlab{a}},
  \apjl, 595, L63

\bibitem[{{De Pontieu} {et~al.}(2004){De Pontieu}, {Erd{\'e}lyi}, \&
  {James}}]{DePontieu2004}
{De Pontieu}, B., {Erd{\'e}lyi}, R., \& {James}, S.~P. 2004, \nat, 430, 536

\bibitem[{{De Pontieu} {et~al.}(2007{\natexlab{a}}){De Pontieu}, {Hansteen},
  {Rouppe van der Voort}, {van Noort}, \& {Carlsson}}]{DePontieu2007}
{De Pontieu}, B., {Hansteen}, V.~H., {Rouppe van der Voort}, L., {van Noort},
  M., \& {Carlsson}, M. 2007{\natexlab{a}}, \apj, 655, 624

\bibitem[{{De Pontieu} {et~al.}(2007{\natexlab{b}}){De Pontieu}, {McIntosh},
  {Hansteen}, {Carlsson}, {Schrijver}, {Tarbell}, {Title}, {Shine}, {Suematsu},
  {Tsuneta}, {Katsukawa}, {Ichimoto}, {Shimizu}, \& {Nagata}}]{DePontieu2007b}
{De Pontieu}, B., {McIntosh}, S., {Hansteen}, V.~H., {Carlsson}, M.,
  {Schrijver}, C.~J., {Tarbell}, T.~D., {Title}, A.~M., {Shine}, R.~A.,
  {Suematsu}, Y., {Tsuneta}, S., {Katsukawa}, Y., {Ichimoto}, K., {Shimizu},
  T., \& {Nagata}, S. 2007{\natexlab{b}}, \pasj, 59, 655

\bibitem[{{De Pontieu} {et~al.}(2003{\natexlab{b}}){De Pontieu}, {Tarbell}, \&
  {Erd{\'e}lyi}}]{DePontieu2003}
{De Pontieu}, B., {Tarbell}, T., \& {Erd{\'e}lyi}, R. 2003{\natexlab{b}}, \apj,
  590, 502

\bibitem[{{Del Zanna} \& {Mason}(2005)}]{DelZanna2005}
{Del Zanna}, G. \& {Mason}, H.~E. 2005, \aap, 433, 731

\bibitem[{{Dere} {et~al.}(1997){Dere}, {Landi}, {Mason}, {Monsignori Fossi}, \&
  {Young}}]{Dere1997}
{Dere}, K.~P., {Landi}, E., {Mason}, H.~E., {Monsignori Fossi}, B.~C., \&
  {Young}, P.~R. 1997, \aaps, 125, 149

\bibitem[{{Domingo} {et~al.}(1995){Domingo}, {Fleck}, \&
  {Poland}}]{Domingo1995}
{Domingo}, V., {Fleck}, B., \& {Poland}, A.~I. 1995, \solphys, 162, 1

\bibitem[{{Doschek} \& {Feldman}(1982)}]{Doschek1982}
{Doschek}, G.~A. \& {Feldman}, U. 1982, \apj, 254, 371

\bibitem[{{Fletcher} \& {de Pontieu}(1999)}]{Fletcher1999}
{Fletcher}, L. \& {de Pontieu}, B. 1999, \apjl, 520, L135

\bibitem[{{Hansteen} {et~al.}(2007){Hansteen}, {Carlsson}, \&
  {Gudiksen}}]{Hansteenetal2007}
{Hansteen}, V.~H., {Carlsson}, M., \& {Gudiksen}, B. 2007, in Astronomical
  Society of the Pacific Conference Series, Vol. 368, The Physics of
  Chromospheric Plasmas, ed. P.~{Heinzel}, I.~{Dorotovi{\v c}}, \& R.~J.
  {Rutten}, 107--+

\bibitem[{{Hansteen} {et~al.}(2006){Hansteen}, {De Pontieu}, {Rouppe van der
  Voort}, {van Noort}, \& {Carlsson}}]{Hansteen2006}
{Hansteen}, V.~H., {De Pontieu}, B., {Rouppe van der Voort}, L., {van Noort},
  M., \& {Carlsson}, M. 2006, Ap. J., 647, L73

\bibitem[{{Heggland} {et~al.}(2007){Heggland}, {De Pontieu}, \&
  {Hansteen}}]{Heggland2007}
{Heggland}, L., {De Pontieu}, B., \& {Hansteen}, V.~H. 2007, \apj, 666, 1277

\bibitem[{{Howard} {et~al.}(2008){Howard}, {Moses}, {Vourlidas}, {Newmark},
  {Socker}, {Plunkett}, {Korendyke}, {Cook}, {Hurley}, {Davila}, {Thompson},
  {St Cyr}, {Mentzell}, {Mehalick}, {Lemen}, {Wuelser}, {Duncan}, {Tarbell},
  {Wolfson}, {Moore}, {Harrison}, {Waltham}, {Lang}, {Davis}, {Eyles},
  {Mapson-Menard}, {Simnett}, {Halain}, {Defise}, {Mazy}, {Rochus}, {Mercier},
  {Ravet}, {Delmotte}, {Auchere}, {Delaboudiniere}, {Bothmer}, {Deutsch},
  {Wang}, {Rich}, {Cooper}, {Stephens}, {Maahs}, {Baugh}, {McMullin}, \&
  {Carter}}]{Howard2008}
{Howard}, R.~A., {Moses}, J.~D., {Vourlidas}, A., {Newmark}, J.~S., {Socker},
  D.~G., {Plunkett}, S.~P., {Korendyke}, C.~M., {Cook}, J.~W., {Hurley}, A.,
  {Davila}, J.~M., {Thompson}, W.~T., {St Cyr}, O.~C., {Mentzell}, E.,
  {Mehalick}, K., {Lemen}, J.~R., {Wuelser}, J.~P., {Duncan}, D.~W., {Tarbell},
  T.~D., {Wolfson}, C.~J., {Moore}, A., {Harrison}, R.~A., {Waltham}, N.~R.,
  {Lang}, J., {Davis}, C.~J., {Eyles}, C.~J., {Mapson-Menard}, H., {Simnett},
  G.~M., {Halain}, J.~P., {Defise}, J.~M., {Mazy}, E., {Rochus}, P., {Mercier},
  R., {Ravet}, M.~F., {Delmotte}, F., {Auchere}, F., {Delaboudiniere}, J.~P.,
  {Bothmer}, V., {Deutsch}, W., {Wang}, D., {Rich}, N., {Cooper}, S.,
  {Stephens}, V., {Maahs}, G., {Baugh}, R., {McMullin}, D., \& {Carter}, T.
  2008, Space Science Reviews, 136, 67

\bibitem[{{Judge} {et~al.}(1995){Judge}, {Woods}, {Brekke}, \&
  {Rottman}}]{Judge1995}
{Judge}, P.~G., {Woods}, T.~N., {Brekke}, P., \& {Rottman}, G.~J. 1995, \apjl,
  455, L85+

\bibitem[{{Kaiser} {et~al.}(2008){Kaiser}, {Kucera}, {Davila}, {St.~Cyr},
  {Guhathakurta}, \& {Christian}}]{Kaiser2008}
{Kaiser}, M.~L., {Kucera}, T.~A., {Davila}, J.~M., {St.~Cyr}, O.~C.,
  {Guhathakurta}, M., \& {Christian}, E. 2008, Space Science Reviews, 136, 5

\bibitem[{{Keenan} {et~al.}(1990){Keenan}, {Tayal}, \& {Henry}}]{Keenan1990}
{Keenan}, F.~P., {Tayal}, S.~S., \& {Henry}, R.~J.~W. 1990, \solphys, 125, 61

\bibitem[{{Klimchuk}(2006)}]{Klimchuk2006}
{Klimchuk}, J.~A. 2006, \solphys, 234, 41

\bibitem[{{Kosugi} {et~al.}(2007){Kosugi}, {Matsuzaki}, {Sakao}, {Shimizu},
  {Sone}, {Tachikawa}, {Hashimoto}, {Minesugi}, {Ohnishi}, {Yamada}, {Tsuneta},
  {Hara}, {Ichimoto}, {Suematsu}, {Shimojo}, {Watanabe}, {Shimada}, {Davis},
  {Hill}, {Owens}, {Title}, {Culhane}, {Harra}, {Doschek}, \&
  {Golub}}]{Kosugi2007}
{Kosugi}, T., {Matsuzaki}, K., {Sakao}, T., {Shimizu}, T., {Sone}, Y.,
  {Tachikawa}, S., {Hashimoto}, T., {Minesugi}, K., {Ohnishi}, A., {Yamada},
  T., {Tsuneta}, S., {Hara}, H., {Ichimoto}, K., {Suematsu}, Y., {Shimojo}, M.,
  {Watanabe}, T., {Shimada}, S., {Davis}, J.~M., {Hill}, L.~D., {Owens}, J.~K.,
  {Title}, A.~M., {Culhane}, J.~L., {Harra}, L.~K., {Doschek}, G.~A., \&
  {Golub}, L. 2007, \solphys, 243, 3

\bibitem[{{Landi} {et~al.}(2006){Landi}, {Del Zanna}, {Young}, {Dere}, {Mason},
  \& {Landini}}]{Landi2006}
{Landi}, E., {Del Zanna}, G., {Young}, P.~R., {Dere}, K.~P., {Mason}, H.~E., \&
  {Landini}, M. 2006, \apjs, 162, 261

\bibitem[{Mariska(1992)}]{Mariska1992}
Mariska, J.~T. 1992, The Solar Transition Region (Cambridge: Cambridge
  University Press)

\bibitem[{{Martens} {et~al.}(2000){Martens}, {Kankelborg}, \&
  {Berger}}]{Martens2000}
{Martens}, P.~C.~H., {Kankelborg}, C.~C., \& {Berger}, T.~E. 2000, \apj, 537,
  471

\bibitem[{{Mazzotta} {et~al.}(1998){Mazzotta}, {Mazzitelli}, {Colafrancesco},
  \& {Vittorio}}]{Mazzotta1998}
{Mazzotta}, P., {Mazzitelli}, G., {Colafrancesco}, S., \& {Vittorio}, N. 1998,
  \aaps, 133, 403

\bibitem[{{McIntosh} {et~al.}(2007){McIntosh}, {Davey}, {Hassler}, {Armstrong},
  {Curdt}, {Wilhelm}, \& {Lin}}]{McIntosh2007}
{McIntosh}, S.~W., {Davey}, A.~R., {Hassler}, D.~M., {Armstrong}, J.~D.,
  {Curdt}, W., {Wilhelm}, K., \& {Lin}, G. 2007, \apj, 654, 650

\bibitem[{{Patsourakos} \& {Klimchuk}(2008)}]{Patsourakos2008}
{Patsourakos}, S. \& {Klimchuk}, J.~A. 2008, \apj, 689, 1406

\bibitem[{{Rouppe van der Voort} {et~al.}(2007){Rouppe van der Voort}, {De
  Pontieu}, {Hansteen}, {Carlsson}, \& {van Noort}}]{Rouppe2007}
{Rouppe van der Voort}, L.~H.~M., {De Pontieu}, B., {Hansteen}, V.~H.,
  {Carlsson}, M., \& {van Noort}, M. 2007, Ap. J. l., 660, L169

\bibitem[{{Rutten}(2006)}]{Rutten2006}
{Rutten}, R.~J. 2006, in Astronomical Society of the Pacific Conference Series,
  Vol. 354, Solar MHD Theory and Observations: A High Spatial Resolution
  Perspective, ed. J.~{Leibacher}, R.~F. {Stein}, \& H.~{Uitenbroek}, 276--+

\bibitem[{{Schmahl} \& {Orrall}(1979)}]{Schmahl1979}
{Schmahl}, E.~J. \& {Orrall}, F.~Q. 1979, \apjl, 231, L41

\bibitem[{{Schrijver} {et~al.}(2004){Schrijver}, {Sandman}, {Aschwanden}, \&
  {DeRosa}}]{Schrijver2004}
{Schrijver}, C.~J., {Sandman}, A.~W., {Aschwanden}, M.~J., \& {DeRosa}, M.~L.
  2004, \apj, 615, 512

\bibitem[{{Spadaro} {et~al.}(1996){Spadaro}, {Lanza}, \&
  {Antiochos}}]{Spadaro1996}
{Spadaro}, D., {Lanza}, A.~F., \& {Antiochos}, S.~K. 1996, \apj, 462, 1011

\bibitem[{{Warren} \& {Winebarger}(2006)}]{Warren2006}
{Warren}, H.~P. \& {Winebarger}, A.~R. 2006, \apj, 645, 711

\bibitem[{{Warren} {et~al.}(2008){Warren}, {Winebarger}, {Mariska}, {Doschek},
  \& {Hara}}]{Warren2008}
{Warren}, H.~P., {Winebarger}, A.~R., {Mariska}, J.~T., {Doschek}, G.~A., \&
  {Hara}, H. 2008, \apj, 677, 1395

\bibitem[{{Wilhelm} {et~al.}(1995){Wilhelm}, {Curdt}, {Marsch}, {Sch{\"u}hle},
  {Lemaire}, {Gabriel}, {Vial}, {Grewing}, {Huber}, {Jordan}, {Poland},
  {Thomas}, {K{\"u}hne}, {Timothy}, {Hassler}, \& {Siegmund}}]{Wilhelm1995}
{Wilhelm}, K., {Curdt}, W., {Marsch}, E., {Sch{\"u}hle}, U., {Lemaire}, P.,
  {Gabriel}, A., {Vial}, J.-C., {Grewing}, M., {Huber}, M.~C.~E., {Jordan},
  S.~D., {Poland}, A.~I., {Thomas}, R.~J., {K{\"u}hne}, M., {Timothy}, J.~G.,
  {Hassler}, D.~M., \& {Siegmund}, O.~H.~W. 1995, \solphys, 162, 189

\bibitem[{{Winebarger} {et~al.}(2008){Winebarger}, {Warren}, \&
  {Falconer}}]{Winebarger2008}
{Winebarger}, A.~R., {Warren}, H.~P., \& {Falconer}, D.~A. 2008, \apj, 676, 672

\bibitem[{{Young} {et~al.}(2009){Young}, {Watanabe}, {Hara}, \&
  {Mariska}}]{Young2009}
{Young}, P.~R., {Watanabe}, T., {Hara}, H., \& {Mariska}, J.~T. 2009, \aap,
  495, 587

\end{thebibliography}


\end{document}